# Resonant, broadband and highly efficient optical frequency conversion in semiconductor nanowire gratings at visible and UV wavelengths

**M. Scalora[1], J. Trull[2], C. Cojocaru[2], M.A. Vincenti[3,6], L. Carletti[4], D. de Ceglia[4], N. Akozbek[5], C. De Angelis[3,6]**

*1 Charles M. Bowden Research Center, CCDEVCOM AVMC, Redstone Arsenal, AL 35898-5000 - U.S.A.*

*2 Physics Department, Universitat Politècnica de Catalunya, Rambla Sant Nebridi 22, Terrassa, Barcelona, Spain*

*3 Department of Information Engineering – University of Brescia, Via Branze 38 25123 Brescia, Italy*

*4 Department of Information Engineering – University of Padova, Via Gradenigo 6/a 35131 Padova, Italy*

*5 AEgis Technologies Inc., 401 Jan Davis Dr. 35806, Huntsville, AL U.S.A.*

*6 National Institute of Optics (INO-CNR), Via Branze 45, Brescia 25123, Italy*

*michael.scalora.civ@mail.mil*

**Abstract:** Using a hydrodynamic approach we examine bulk- and surface-induced second and third harmonic generation from semiconductor nanowire gratings having a resonant nonlinearity in the absorption region. We demonstrate resonant, broadband and highly efficient optical frequency conversion: contrary to conventional wisdom, we show that harmonic generation can take full advantage of resonant nonlinearities in a spectral range where nonlinear optical coefficients are boosted well beyond what is achievable in the transparent, long-wavelength, non-resonant regime. Using femtosecond pulses with approximately 500 MW/cm$^2$ peak power density, we predict third harmonic conversion efficiencies of approximately 1% in a silicon nanowire array, at nearly any desired UV or visible wavelength, including the range of negative dielectric constant. We also predict surface second harmonic conversion efficiencies of order 0.01%, depending on the electronic effective mass, bistable behavior of the signals as a result of a reshaped resonance, and the onset fifth order nonlinear effects. These remarkable findings, arising from the combined effects of nonlinear resonance dispersion, field localization, and phase-locking, could significantly extend the operational spectral bandwidth of silicon photonics, and strongly suggest that neither linear absorption nor skin depth should be motivating factors to exclude either semiconductors or metals from the list of useful or practical nonlinear materials in any spectral range.

The advent of metamaterials and the search for efficient nonlinear frequency converters have mediated the growth of theoretical and experimental studies of second and third harmonic generation (SHG and THG) from all types of metallic, semiconductor, and all-dielectric nanostructures [1]. High harmonic generation is an important process in achieving coherent light sources at shorter wavelengths [2]. Most efforts have focused primarily on semiconductors and all-dielectric structures to avoid losses that are thought to be inherent to metallic systems, leading researchers not only to set aside metals, but also to concentrate on the transparency ranges of dielectrics and semiconductors. The study of nonlinear optics at the nanoscale generally requires modification of the material equations of motion to account for nonlocal effects, quantum tunneling, electron spill-out and screening effects, and surface phenomena [3]. Yet, even in transparent materials most studies are concerned with the scrutiny of bulk nonlinearities neglecting their microscopic origins, while little or no attention is given to simultaneously accounting for: (i) the dynamics of linear and nonlinear surface phenomena in the femtosecond regime; (ii) concurrent inclusion of second, third and possibly higher order processes; (iii) competing nonlinearities; (iv) pump depletion and the possibility of down-conversion; (v) optical processes at wavelengths below the band edge, in the opacity range.

It is generally thought that the wavelength range below the band edge of ordinary semiconductors is a hostile environment for light, so much so that, for example, clever strategies are devised in order to avoid the propagation of the generated harmonic signal inside absorbing material [4]. Indeed, broad absorption resonances characterize most semiconductors in that regime, with metallic [$\mathrm{Re}(\varepsilon) < 0$] response below 300nm [5], and an epsilon-near-zero (ENZ) crossing near 100nm. Similar circumstances occur in materials that display a Lorentzian response [6], suggesting that this ENZ crossing may be potentially exploitable for high-harmonic generation [7]. That notwithstanding, it has been predicted and experimentally demonstrated that SH and TH signals can be generated in the UV range, and propagate through GaAs [8] and GaP [9] substrates that are hundreds of microns in thickness. The harmonic conversion process of crossing a single surface is inherently inefficient due to the lack of phase matching, but it can be rendered somewhat effective in semiconductor cavities sandwiched between Bragg reflectors, which have been demonstrated to yield relatively large conversion efficiencies in the visible range [10].

This unique phenomenology is pivotal to first understand and then control nonlinear optical phenomena in the opaque regime of all semiconductors, and is due to the onset of a phase locking mechanism [11] that allows a pump signal tuned in the transparency range to impress its phase and absorption properties to the harmonic signals, triggering the inhibition of linear absorption for each generated harmonic. In the undepleted pump approximation, the solutions of Maxwell's equations in nonlinear materials that display SHG or THG have two parts [12]: one solves the homogeneous equation; the other is a particular solution of the inhomogeneous equation driven by the nonlinear polarization term induced in the nonlinear medium by the pump field. The inhomogeneous harmonic signal refracts at the same angle and travels at the same velocity as the pump; the homogeneous component walks off from the pump, refracts and travels according to the values one expects from material dispersion at the harmonic frequency. These dynamics have been clearly experimentally demonstrated in a cm-long, transparent lithium niobate crystal: a first SH spot was observed at the same exit angle as the pump, while a second SH spot was observed at the refraction angle one expects from material dispersion at the SH wavelength [13]. The dynamics demonstrated in references [8] and [9] extended the notion of phase locking to the imaginary part of the refractive index, thus opening up the entire opaque range of semiconductors to investigation: if the pump is tuned to the transparent or semi-transparent range and the harmonic signals are tuned at wavelengths below the band edge, the inhomogeneous signals survive, can co-propagate over long distances, and can resonate with the pump yielding unusually large SHG and THG conversion efficiencies [10,14]. Our results below thus suggest that the generated harmonic signals also appear to exploit the large, resonant nonlinearities that characterize the material in the opacity range, leading to unprecedented conversion efficiencies at visible and UV wavelengths.

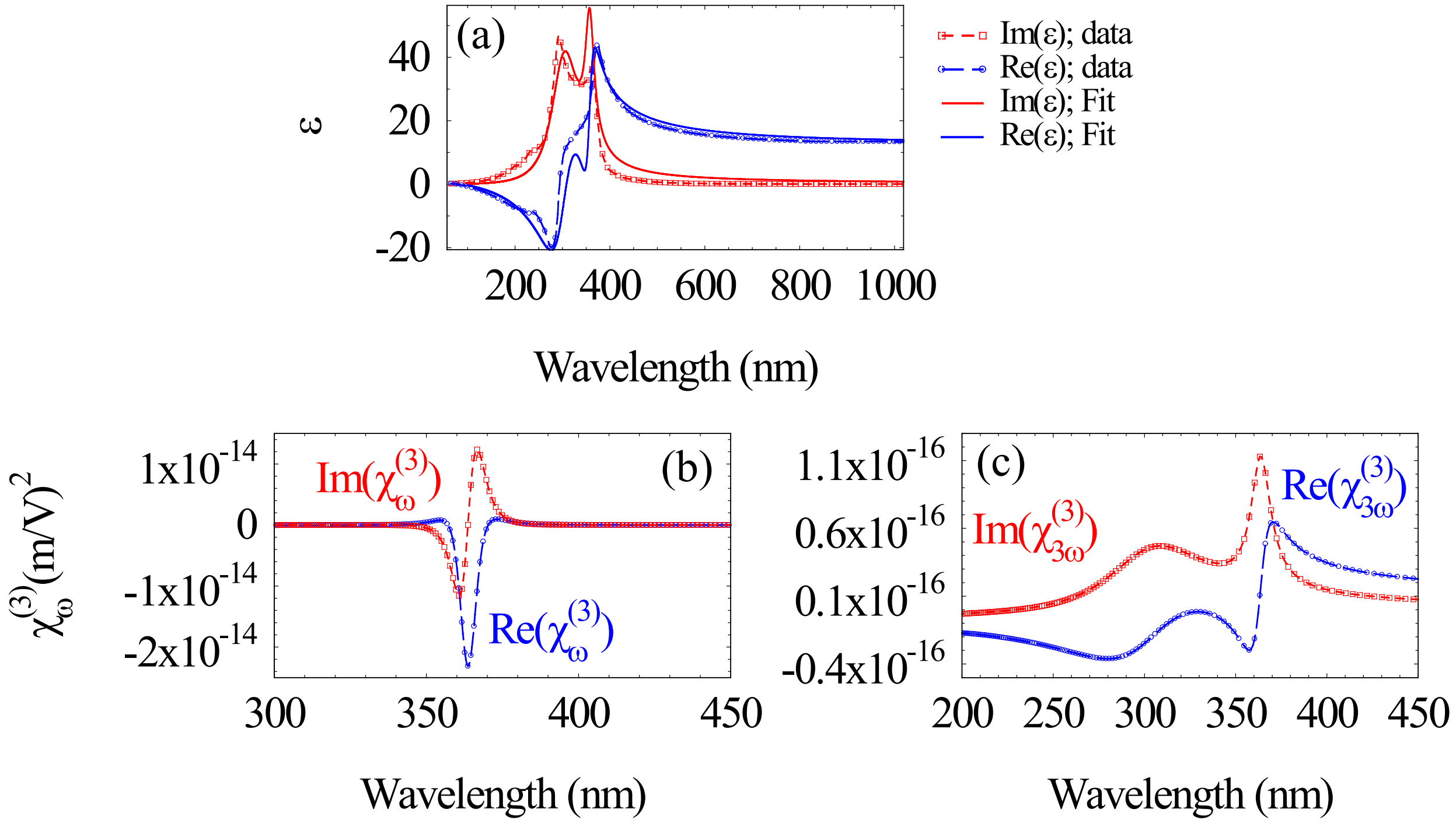

**Fig.1**: (a) Bulk silicon data found in reference [5] (curves with markers) and fit for real and imaginary parts of dielectric permittivity (solid curves); nonlinear susceptibility calculated using linear oscillator parameters: (b) Kerr Coefficient (self-phase modulation and nonlinear pump absorption, Eq.(1)). At 1064nm $\chi_{\omega}^{(3)} \approx 3\times10^{-19}(\mathrm{m/V})^2$; (c) third harmonic generation coefficient, Eq.(2).

In Fig.1a we plot the dielectric constant of undoped crystalline silicon, as reported in Palik's handbook of optical constants (as indicated in the legend.) Two closely-spaced resonances are found in the 300nm to 400nm range. The solid curves are obtained using two separate Lorentzian resonances, having distinct damping coefficients, resonance and plasma frequencies, as illustrated in what follows. Absorption decreases rapidly above 800nm. In Figs.1b and 1c we plot the third order nonlinear dispersions associated with the pump (b) and TH (c) signals using the linear oscillator parameters used to fit the data [15, 16]. The expressions that generate Figs.1b and 1c may be derived using a perturbative approach and may be written as follows [17]:

$$\chi_{\omega}^{(3)} = \frac{n_1\beta e^4}{m^3(\omega_{01}^2-\omega^2+i\gamma_1\omega)^3(\omega_{01}^2-\omega^2-i\gamma_1\omega)} + \frac{n_2\beta e^4}{m^3(\omega_{02}^2-\omega^2+i\gamma_2\omega)^3(\omega_{02}^2-\omega^2-i\gamma_2\omega)}, \quad (1)$$

and

$$\chi_{3\omega}^{(3)} = \frac{n_1\beta e^4}{m^3(\omega_{01}^2-9\omega^2-3i\gamma_1\omega)(\omega_{01}^2-\omega^2-i\gamma_1\omega)^3} + \frac{n_2\beta e^4}{m^3(\omega_{02}^2-9\omega^2-3i\gamma_2\omega)(\omega_{02}^2-\omega^2-i\gamma_2\omega)^3}, \quad (2)$$

where $m$ is electron's effective mass and $\beta$ is the third order nonlinear oscillator parameter described below. The simultaneous inclusion of linear and nonlinear dispersive properties is pivotal if one is to accurately depict second and third harmonic generation, unless all signals are tuned far into the IR range, tuned far off resonance, where dispersion is flat. For crystalline, centrosymmetric silicon, the second harmonic signal is generated entirely by the surface and by magnetic Lorentz contributions.

The material model that we use to derive Eqs.(1-2) may be summarized by two coupled equations that describe the two resonances reported in Fig.1a, which in the linear regime yield back the local dielectric functions. The oscillators (dipoles) generally experience linear and nonlinear internal restoring forces, damping, and external forces in the form of the applied electromagnetic field that couples the equations. In turn, the fields are expanded near the surface, where they vary rapidly, yielding hydrodynamic-like equations [17, 18] that contain the spatial derivatives of the electric field (or polarizations) that drive most of the second harmonic signal generated at the surface:

$$\ddot{\mathbf{P}}_1 + \gamma_1 \dot{\mathbf{P}}_1 + \omega_{01}^2 \mathbf{P}_1 + \alpha_1 \mathbf{P}_1 \mathbf{P}_1 - \beta_1 (\mathbf{P}_1 \bullet \mathbf{P}_1) \mathbf{P}_1 + \delta_1 (\mathbf{P}_1 \bullet \mathbf{P}_1)^2 \mathbf{P}_1 = \frac{n_1 e^2}{m_1} \mathbf{E} + \frac{e}{m_1} (\mathbf{P}_1 \bullet \nabla) \mathbf{E} + \frac{e}{m_1 c} \dot{\mathbf{P}}_1 \times \mathbf{H}$$

$$\ddot{\mathbf{P}}_2 + \gamma_2 \dot{\mathbf{P}}_2 + \omega_{02}^2 \mathbf{P}_2 + \alpha_2 \mathbf{P}_2 \mathbf{P}_2 - \beta_2 (\mathbf{P}_2 \bullet \mathbf{P}_2) \mathbf{P}_2 + \delta_2 (\mathbf{P}_2 \bullet \mathbf{P}_2)^2 \mathbf{P}_2 = \frac{n_2 e^2}{m_2} \mathbf{E} + \frac{e}{m_2} (\mathbf{P}_2 \bullet \nabla) \mathbf{E} + \frac{e}{m_2 c} \dot{\mathbf{P}}_2 \times \mathbf{H}$$

. (3)

We have neglected higher order contributions, and for now we also neglect second order bulk nonlinearities ($\alpha_1 = \alpha_2 = 0$,) which are reintroduced for semiconductors like GaAs and GaP [9, 17, 18]. In view of our results for silicon, we maintain third and fifth order nonlinearities and purposely neglect nonlinearities of the fourth order.

Notwithstanding the apparent complexities of Eqs.(3), the derivation of these equations assumes that for each oscillator all damping coefficients ($\gamma_l$), resonance ($\omega_{0l}$) and plasma ($\omega_{pl}^2 = \frac{4\pi n_l e^2}{m_l}$) frequencies, are identical in all spatial directions, where the subscript $l$ refers to either oscillator 1 or 2. This is probably an oversimplification. For example, the two oscillator species highlighted in Eqs.(3) and Fig.1a have different plasma frequencies. One may then assume that either the densities or the effective masses, or both, are different in order to satisfy this condition. From a microscopic point of view, the densities $n_1$ and $n_2$ correspond to the excitation of different orbitals within the valence band. It is possible that any given orbital may contribute more electrons at one resonance compared to another. However, the increase or decrease of the number densities may not necessarily be commensurate with the respective effective masses, which may also be different for different orbitals and different spatial directions, depending on the spring constants and the orbitals' symmetries. These uncertainties ultimately delineate the limits of the classical model. In practice, then, once the dielectric function has been fitted with adequate Lorentzian functions, the magnitude of the effective masses are the only free parameters that determine the conversion efficiencies of surface-generated harmonics. Concurrently, the magnitude of the nonlinear parameters $\alpha_l$, $\beta_l$ and $\delta_l$ are such that $|\alpha_l| \approx \frac{\omega_{0l}^2}{L}$, $|\beta_l| \approx \frac{\omega_{0l}^2}{L^2}$ and $|\delta_l| \approx \frac{\omega_{0l}^2}{L^4}$ [15], where $L$ is the approximate lattice constant of the given material, which typically is of order 3-5Å. The relationship between these coefficients and the respective nonlinear susceptibilities may be derived using the nonlinear Lorentz oscillator model [16]. For instance, the connection between $\beta$ and $\chi^{(3)}_{\omega,3\omega}$ is described by Eqs.1-2. Similar expressions may be derived for each order of dispersion. As a result, keeping in mind the basic limitations and degrees of freedom of the classical model, Eqs.(3) are quite general and may be adapted to any non-conductive material, provided sufficient care is exercised in the expansion of the powers of the polarizations, and that the tensor properties of the nonlinear susceptibilities of the material are appropriately taken into account.

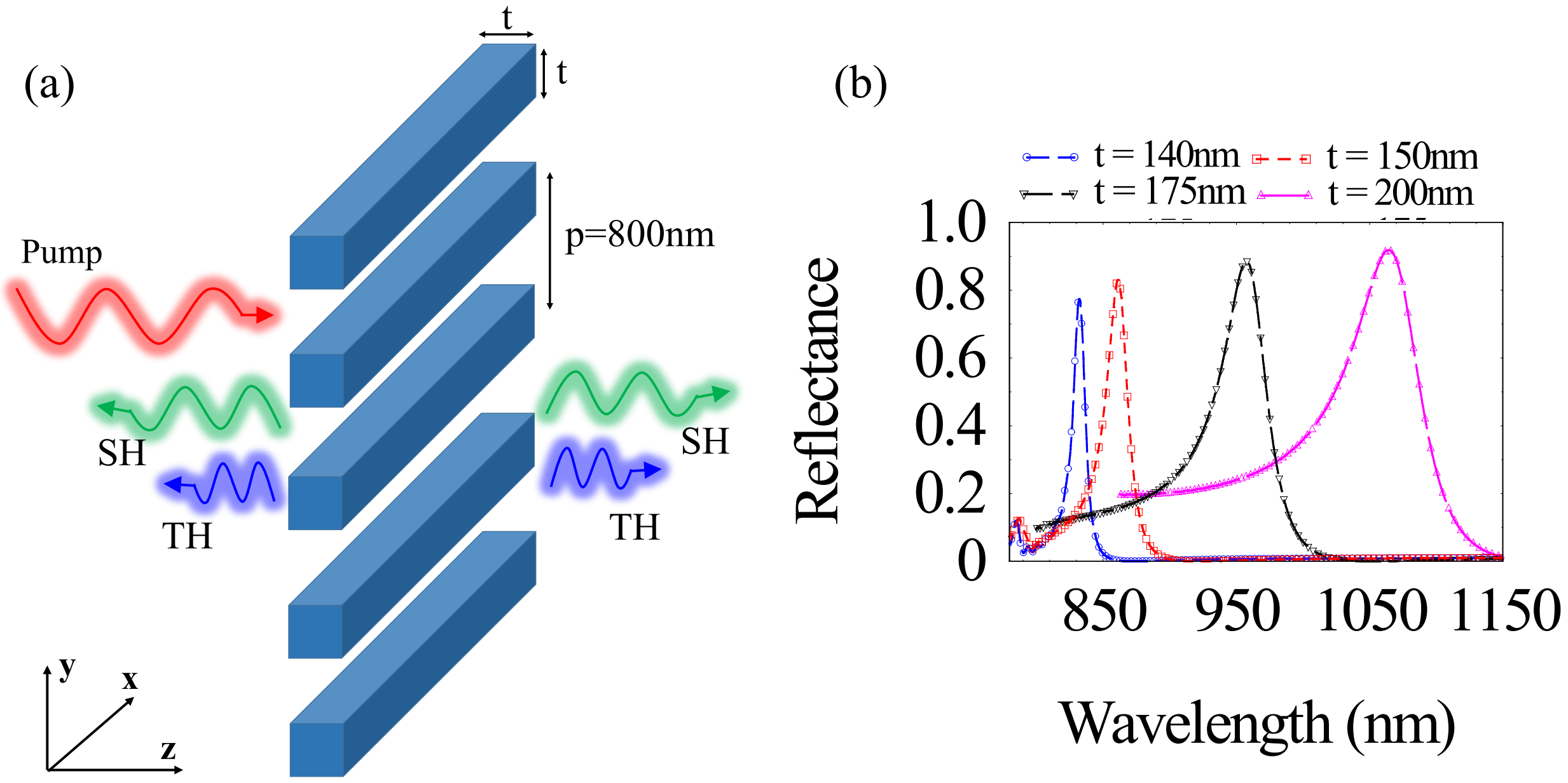

**Fig.2:** (a) a pump beam or pulse is incident from the left along the *z*-direction on a silicon nanowire array of cross section $t \times t$ and periodicity *p*=800nm. An incident TM-polarized field has a single magnetic field component along *x* and electric field components along *y* and *z*. Conversely, an incident TE-polarized field has a single electric field component along *x* and magnetic field components along *y* and *z*. (b) Linear reflectance that shows a tunable band structure depending on the cross section of the nanowire for a TM-polarized field at normal incidence.

We now consider the optical response of an array composed of silicon nanowires at normal incidence, as illustrated in Fig. 2a, having cross section $t \times t$ and periodicity *p*=800nm. As shown in Fig. 2b, smaller cross sections blueshift and narrow the resonance. In contrast, smaller periodicities redshift the resonance (not shown). Therefore, using a variety of optical materials one may control almost at will the position of the resonance by varying these two parameters simultaneously, as we will see later for GaP and GaAs. In Figs. 3a and 3b we summarize the predictions for the total TM-polarized SH signal generated by the surface and the magnetic Lorentz terms – Fig. 3a – and total TH signal – Fig. 3b – originating in the bulk of Si nanowires having t=150nm. The energy is nearly equally split between transmission and reflection, with remarkable conversion efficiencies of nearly 1% for the TH and 0.01% for the SH signal. Figs. 3a and 3b contain four curves, which correspond to a particular transform-limited, Gaussian pulse duration (80fs to 640fs, as indicated) and are obtained by sweeping the carrier wavelength from ~830nm to ~920nm. The longest pulse ultimately resolves bistable behavior in both SH and TH signals. That is to say, the output SH and TH conversion efficiencies display a drop as the carrier wavelength of the 640fs pulse is tuned from 890nm to 891nm. Increasing pulse duration is equivalent to probing the structure with narrower pulse bandwidths, which tend to resolve the resonance and in turn couple more strongly with the structure, leading to larger local field intensities and lower nonlinear thresholds. In Fig. 3c we highlight the importance of including fifth order nonlinearities in the dynamics of oscillators that describe silicon. A non-zero $\delta_l$ coefficient, equivalent to the excitation of $\chi^{(5)}$ and having sign opposite to that of $\beta_l$, tends to decrease the overall nonlinearity, consequently slightly increasing nonlinear thresholds compared to having only $\chi^{(3)}$ (non-zero $\beta_l$) contributions. Finally in Fig. 3(d) we depict the typical TM-polarized pump field intensities inside the nanowires at each of the resonances displayed in Fig. 2b, which display amplification factors that range from ~20 for the electric field intensity to ~600 for the magnetic field intensity. This is a clear indication that the energy velocity of the incident pulse is reduced between two and three orders of magnitude relative to the speed of light in vacuum.

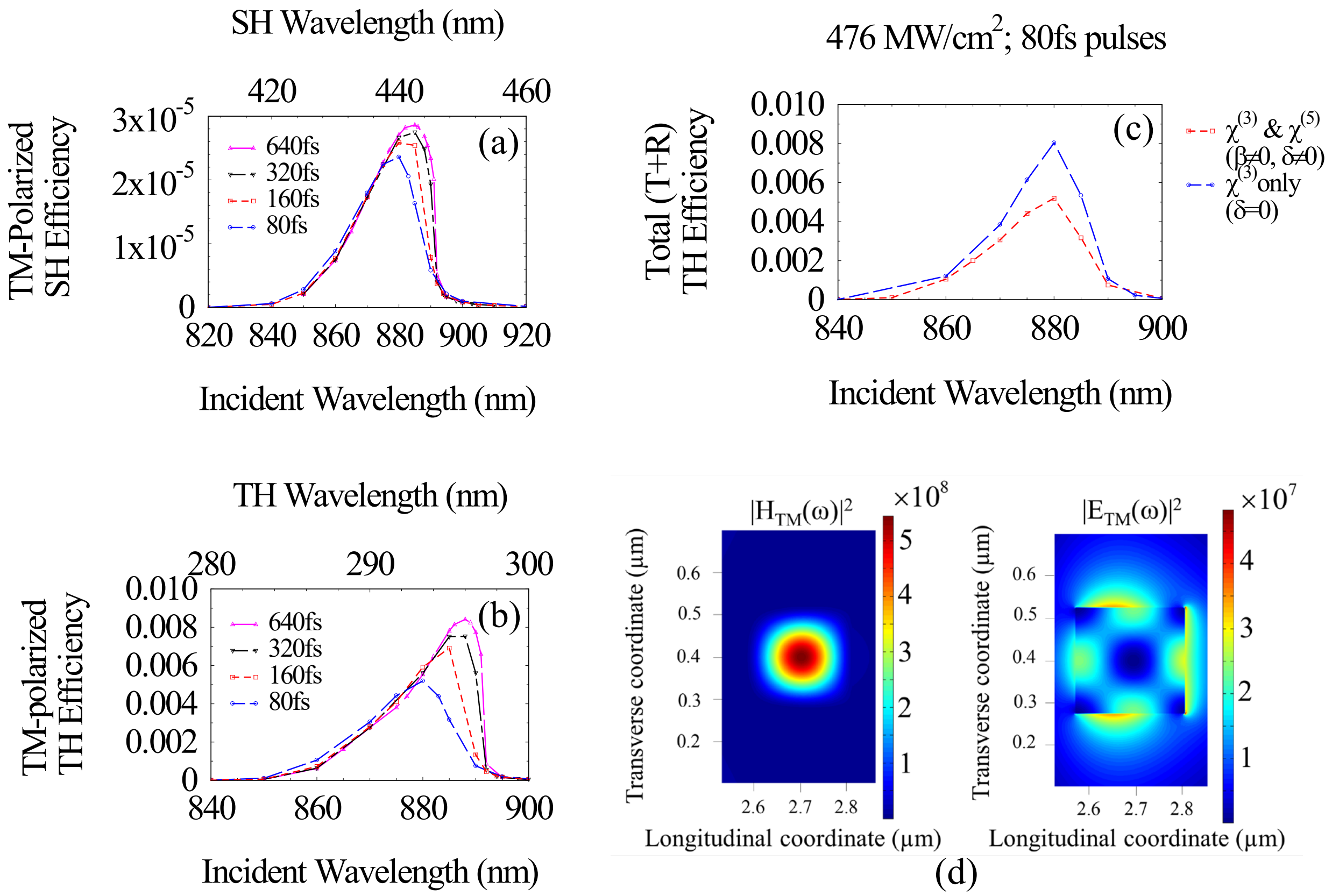

**Fig.3:** Total (a) SH and (b) TH conversion efficiency vs. incident carrier wavelength as a function of incident pulse width for a grating having t=150nm. The incident field is TM-polarized and has peak power density ~500MW/cm$^2$. Pulses of longer duration give way to more intense local fields inside each nanowire compared to shorter pulses, thus tending to reduce nonlinear threshold. Both SH and TH signals display optical bistability near 890nm. (c) The fifth order nonlinearity affects the dynamics by lowering TH conversion efficiencies and thus increases nonlinear thresholds. (d) Typical TM-polarized magnetic ($|H_{TM}|^2$) and electric ($|E_{TM}|^2$) pump field modes recorded on resonance when the peak of the pulse reaches the grating. The magnetic field forms a strongly localized Gaussian; the electric field circulates near and envelopes the walls of the structure. This field delocalization reduces the magnitude of the Poynting vector leading to slow light. The effective mass of the electron is chosen as $m$=0.015$m_e$. This value directly affects surface SHG, and indirectly determines third order nonlinear dispersion, e.g. Eqs.(1-2).

Now recall that if the pump were to cross a single interface harmonic generation would be rather inefficient [8, 9, 10, 13]. The remarkable conversion efficiencies that we predict, ~1% for THG and ~0.01% for surface SHG, do not come directly as a result of electric field intensity amplification inside the cavity, which is a modest factor of 20. Rather, the harmonics are able to extract energy from the pump quite efficiently as a result of the repeated reflections associated with circulation of the pump field around the nanowire, and the fact that the harmonics appear to effectively lock onto the large nonlinear coefficients that the pump experiences as a result of a resonant nonlinearity, as outlined in Fig.1. We note that the conversion efficiencies are sensitive to the effective mass of the electron (surface SHG) and to the magnitude of $\beta_l$. These parameters may be extracted from experimental observations of surface SHG and bulk generated TH signal from a simple substrate of the material in question [8-10].

We now briefly describe THG results for GaAs and GaP nanowire arrays. The linear dielectric response of these materials closely resembles that of Si represented in Fig.1 [5], except for the fact the resonances are further apart and the magnitudes of Re (ε) and Im (ε) are roughly a factor of two smaller, with consequently smaller nonlinear response. As mentioned above, the method encapsulated in Eqs.(3) employs a nonlinear Lorentz oscillator model that naturally captures both linear and nonlinear material dispersions, while simultaneously accounting for surface and magnetic nonlinearities. In general, the third order bulk nonlinearity described in Eqs.(3) may be written as, and $P_i^{NL^{(3)}} = \sum_{j=1,3,} \sum_{k=1,3,} \sum_{l=1,3,} \beta_{i,j,k,l} P_j P_k P_l$. For isotropic GaAs and GaP this expression generates the following, simplified vector components:

$$\begin{pmatrix} P_x^{NL^{(3)}} \\ P_y^{NL^{(3)}} \\ P_z^{NL^{(3)}} \end{pmatrix} = \beta \begin{pmatrix} \left(P_x^2 + P_y^2 + P_z^2\right) P_x \\ \left(P_x^2 + P_y^2 + P_z^2\right) P_y \\ \left(P_x^2 + P_y^2 + P_z^2\right) P_z \end{pmatrix} \quad . \qquad (4)$$

For simplicity we have consolidated the constants into a single coefficient, assumed that these coefficients are identical for the two oscillator species we have identified in Eqs.(3), and report only THG results to compare directly with the results above.

In Fig.4 we summarize the linear reflection profiles –Fig. 4a– and relative THG conversion efficiencies –Fig. 4b– for two separate GaAs and GaP nanowire gratings having different cross section and periodicity, as indicated in the figure. The size of the nanowires and periodicity of the array can be used to tune the optical band structure near the electronic band edge (~490nm for GaP and ~910nm for GaAs,) in order to minimize the imaginary part of the dielectric constant for the pump field, and thus maintain the requirement that the pump scatter without being absorbed. All things being equal, i.e. effective electron masses, incident peak power density, resonance frequencies and lattice constants, Si outperforms the THG conversion efficiency of both GaP and GaAs by nearly two orders of magnitude. This discrepancy is explained by the fact that a straightforward calculation shows that $\chi_{\omega}^{(3)}$ for GaP and GaAs are nearly one order of magnitude smaller compared to silicon, i.e. Figs.1b and 1c, consistent with two orders of magnitude change in nonlinear TH gain. To further emphasize the point that a resonant nonlinearity is mostly responsible for these remarkable results, we also calculated the conversion efficiencies depicted in the figures above without the benefit of nonlinear Lorentz oscillators, i.e. by simply inserting the third order susceptibility given by:

$$\begin{pmatrix} P_x^{NL^{(3)}} \\ P_y^{NL^{(3)}} \\ P_z^{NL^{(3)}} \end{pmatrix} = \chi^{(3)} \begin{pmatrix} \left(E_x^2 + E_y^2 + E_z^2\right) E_x \\ \left(E_x^2 + E_y^2 + E_z^2\right) E_y \\ \left(E_x^2 + E_y^2 + E_z^2\right) E_z \end{pmatrix} \quad , \qquad (5)$$

using a dispersionless nonlinear coefficient $\chi^{(3)} \approx 3\times10^{-19}(\mathrm{m/V})^2$, estimated at the pump wavelength. Predictably, conversion efficiencies were reduced between two and three orders of magnitude. Therefore, one cannot overstate the importance of simultaneously incorporating nonlinear dispersion, surface and magnetic phenomena in any calculation of this type.

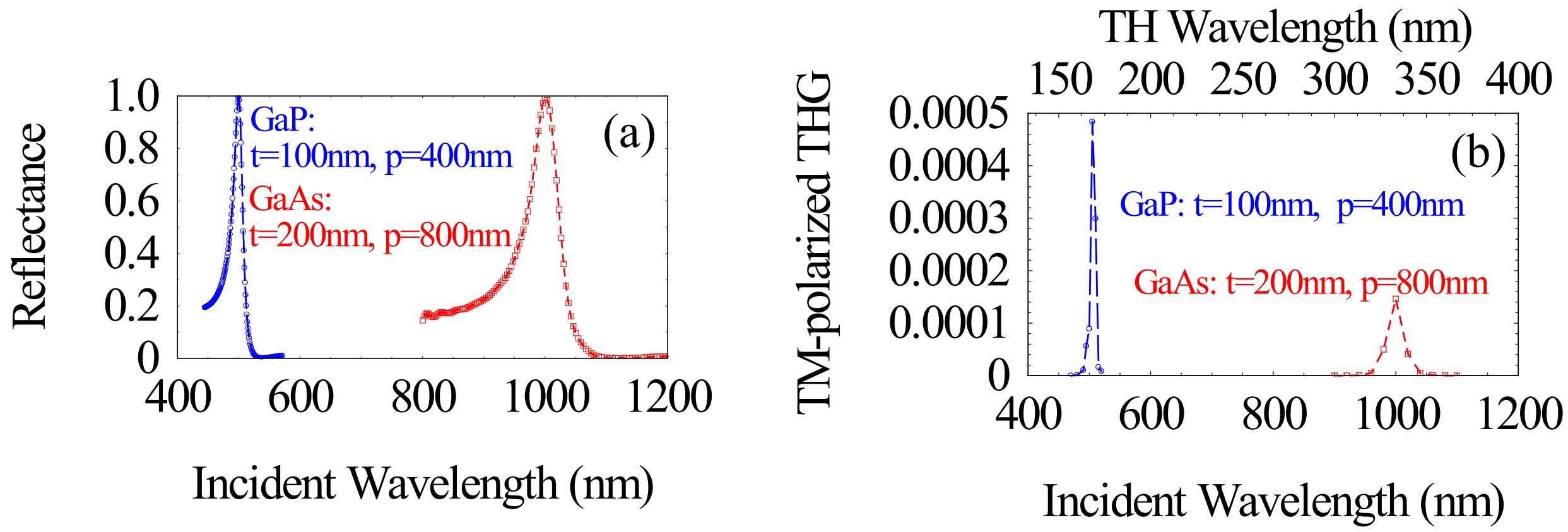

**Fig.4:** (a): Linear reflection for GaP and GaAs grating having different cross sections (t=100nm and t=200nm, respectively) and periodicities (p=400nm and p=800nm, respectively.) (b) THG conversion efficiencies for the gratings described in the left panel. The GaP grating has narrower resonances that correspond to higher cavity quality factors, and yield larger conversion efficiencies compared to the GaAs grating.

In summary, using a hydrodynamic approach we have demonstrated that it is theoretically possible to exploit field localization, resonant nonlinearity, and a phase locking mechanism to obtain nonlinear TH conversion efficiencies of order 1% using a nanowire array composed of ordinary semiconductors like silicon and a peak power density of order 500MW/cm$^2$. These unprecedented conversion efficiencies may vary depending on the precise magnitude of the nonlinear coefficients in Eqs.(3), which depend on resonance frequencies, lattice constants, effective masses, and other geometric form or shape factors, and are possible because the harmonic signals survive in the opaque regime (visible and UV), resonate with the pump, and adopt the large nonlinear second and third nonlinear gain coefficients associated with the pump field. A phase locking mechanism subjects the harmonics to the propagation properties of the pump, including phase and group velocities, lack of absorption and nonlinear optical properties. The wave vector associated with the $j^{\text{th}}$ phase-locked harmonic signal is given by [10] $k_{j\omega} = jk_\omega = j\frac{\omega}{c}n(\omega)$, where the total, complex index of refraction is $n(\omega) = n_{Linear}(\omega) + n_{Nonlinear}(\omega)$, which also contains nonlinear contributions. Using this approach it then becomes possible to exploit ordinary semiconductors to generate harmonics rather efficiently at wavelengths well below 300nm, suggesting the possibility of new coherent sources in the deep UV spectral region.

**Acknowledgements**

MS, NA, JT and CC acknowledge useful discussions with V. Roppo, G. Leo and Z. Coppens. MAV acknowledges financial support from the Rita Levi-Montalcini Italian Ministry of Education and Research. JT and CC acknowledge financial support from RDECOM Grant W911NF-16-1-0563 from the International Technology Center-Atlantic.